\documentclass{caosp306}

\usepackage{graphicx}
\usepackage{siunitx}
\usepackage{xcolor}

\usepackage{natbib}
\articleNo{1}
\pubyear{2020}
\volume{1}
\volnumber{1}
\firstpage{1}
\received{July 31, 2019}
\accepted{August 28, 2019}

\def\BibTeX{{\rm B\kern-.05em{\sc i\kern-.025em b}\kern-.08em
             T\kern-.1667em\lower.7ex\hbox{E}\kern-.125emX}}

\begin{document}

%

\htitle{Spectral Classification of M-dwarf Candidate Stars}
\hauthor{Cabello {\it et al.}}

\title{Independent Study and Spectral Classification of a Sample of Poorly Studied High Proper Motion M-dwarf Candidate Stars}


%
\author{
        Cristina Cabello \inst{1}
        \and 
        G\'eza Cs\"ornyei  \inst{2,} \inst{3} 
        \and 
        Jaroslav Merc \inst{4,}\inst{5}
        \and 
        Ver\'onica Ferreir\'os Lopez \inst{6}
        \and
        Peter Pessev \inst{7,} \inst{8,} \inst{9}
       }

%
          
\institute{
            Departamento de F\'isica de la Tierra y Astrof\'isica. Instituto de F\'isica de Part\'iculas y del Cosmos (IPARCOS), Universidad Complutense de Madrid, E-28040 Madrid, Spain\\ \email{criscabe@ucm.es}
            \and
            Institute of Physics, E\"otv\"os Lor\'and University, P\'azm\'any P\'eter s\'et\'any 1/a, 1117 Budapest, Hungary\\ \email{csornyei.geza@csfk.mta.hu}
            \and
            Konkoly Observatory, Research Centre for Astronomy and Earth Sciences, Konkoly Thege Mikl\'os \'ut 15-17, 1121 Budapest, Hungary
            \and
            Astronomical Institute, Faculty of Mathematics and Physics, Charles University, V Hole\v{s}ovi\v{c}k{\'a}ch 2, 180 00 Prague, Czech Republic\\ \email{jaroslav.merc@gmail.com}
            \and
            Institute of Physics, Faculty of Science, P. J. \v{S}af{\'a}rik University, Park Angelinum 9, 040 01 Ko\v{s}ice, Slovak Republic
            \and
            Physics department, Lancaster University\\ LA1 4YW Lancaster, United Kingdom\\ \email{veronica.ferreiros@gmail.com}
            \and
            Gran Telescopio Canarias (GTC), La Palma, Tenerife, Spain\\ \email{peter.pessev@gtc.iac.es}
            \and
            Instituto de Astrof\'isica de Canarias (IAC), La Laguna, Tenerife, Spain
            \and
            Universidad de La Laguna, Departamento de Astrof\'isica, La Laguna, Tenerife, Spain
          }
\date{March 8, 2003}

\maketitle

\begin{abstract}
We report an independent spectral classification of a sample of poorly studied M-dwarf candidate stars observed with the OSIRIS instrument at GTC. Our project was carried out as an independent test of the spectral classification. It is crucial for the studies of extrasolar planets orbiting M-dwarfs, since properties of the host star are directly related to understanding the planet properties and possible habitability. Understanding of the statistical properties of the dwarf stars is also crucial for the Simple Stellar Population models that play a major role in the modern astrophysics. H$\alpha$ emission was detected in 33\% of the sample with evidence of H$\alpha$ variability in one object.
\end{abstract}

%
\section{Introduction}
Late-type dwarfs are the least massive (M $\sim$ 0.08 - 0.60 M$\odot$) and coolest stars (T$\rm_{eff}$ $\sim$ 2300 - 3800 K) on the main sequence. They are the most populous objects in the Galaxy \citep[up to $\sim$ 70\% of all stars,][]{1997AJ....114..388H}, but their observations are difficult due to their low luminosity (L $\sim$ 0.0002 - 0.08 L$\odot$). Analysis of their physical properties is essential for the characterisation of the population of low-mass stars in the Galaxy. It also has significant impact on the initial mass function (IMF), simple stellar population (SSP) and evolutionary population synthesis (ESP) models. Some red dwarfs are known hosts of extrasollar planets (also of "super-Earth" size).

These stars evolve very slowly (for trillions of years), moreover red dwarfs with the mass less than 0.35 M$\odot$ are fully convective \citep{2009A&A...496..787R} therefore the produced helium is remixed with the material of the star prolonging the time they spend on the main sequence. This is the reason why late-type dwarfs have not reached advanced stages of their evolution yet.

Their spectra are dominated by the absorption molecular bands. Some of them reveal strong magnetic activity (Balmer lines, mainly H$\alpha$ in emission).

\section{Target selection and instrumental setup}

This  work  is  primarily  based  on  optical spectroscopic  data  obtained with the OSIRIS instrument at Gran Telescopio Canarias  (GTC), using Long Slit mode. The configuration was R1000R+GR, covering the wavelength range of 5100 - 10000~\AA. Observations were carried out at the parallactic angle trough the 1 arcsecond long slit. 
The data set was obtained through a queue programme between September 2016 and January 2017 (semester 16B). The basics of the sample selection is relying on the 2MASS color indices \citep{2015MNRAS.446.3878M} in order to avoid contamination by giant stars or galaxies. A limit of proper motion greater of 0.3 arcseconds/year was imposed in order to separate nearby dwarfs from more distant giants. The sample selection was carried out before GAIA DR1 and the proper motions were mostly derived from 2MASS and ALLWISE with a baseline of the observations between 9 and 13 years.
Objects in the range 9 $<$ J $<$ 16 were selected for the observations since brighter objects are most likely already studied and fainter ones were not suitable for observing within relatively short observing blocks even with GTC.

\section{Data reduction}
Each obtained spectral frame has been processed with the basic long slit spectroscopy reduction included in IRAF \citep{1986SPIE..627..733T}. 
By combining the individual spectra in each observing block, significant amount of the random noise introduced by the cosmic rays have been mitigated. The resulting spectra were  sufficient for the type determination, which has been performed with the \texttt{pyhammer} \texttt{python} package \citep{2017AAS...22924035K}. This package determines the spectral type of the object by comparing various spectral templates to the input spectra, then determining the best fitting one with the help of the least square fit method. 

\section{Results}

\begin{figure}[!t]
    \centering
    \includegraphics[width=0.95\linewidth]{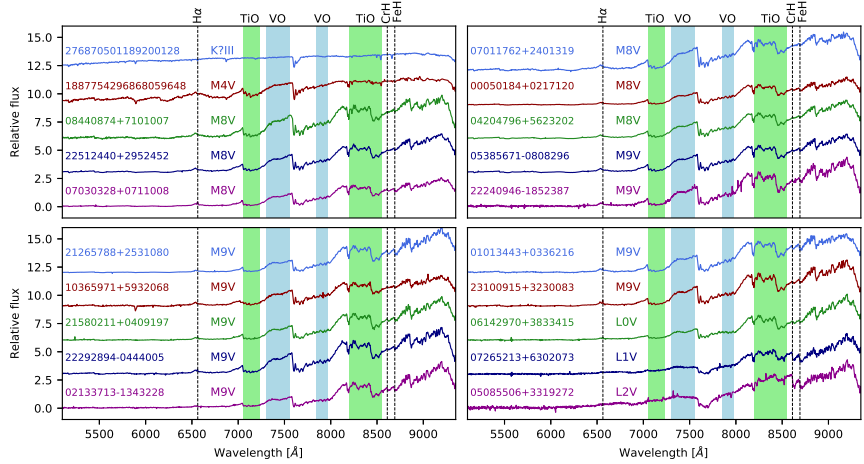}
    \caption{The extracted spectra along with the spectral types determined by \texttt{pyhammer} and their identifiers from 2MASS. For those two objects, which had no 2MASS identifier (top left panel) the Gaia identifier were given.}
    \label{fig:result_spectra}
\end{figure}


We have analysed a total of 20 spectra: 16 corresponds to M-dwarfs, from types M4 to M9, 3 are classified as early types of L-–dwarfs (L0-L2). The object Gaia DR2 276870501189200128 is a K giant star. Figure \ref{fig:result_spectra} shows the spectra for each of our targets\footnote{We should note that the K giant and the M4 dwarf, were not primary targets of the observations and their spectra were recorded together with other dwarf stars only by chance.}. Oxides molecules like TiO $\lambda$7053 \si{\angstrom} or VO $\lambda$7400 \si{\angstrom} bands which dominate the far-optical portions of late M-spectra are replaced by metallic hybrids like FeH $\lambda$8692 \si{\angstrom} and CrH $\lambda$8611 \si{\angstrom} or neutral alkalis (doublets Rb I $\lambda$7800 \si{\angstrom} $\lambda$7948 \si{\angstrom}, Cs I $\lambda$8521 \si{\angstrom} $\lambda$8943\si{\angstrom}, Na I $\lambda$5889 \si{\angstrom} $\lambda$5895 \si{\angstrom}) as the strongest and more significant features in early L-type stars. M-dwarfs can show signals of chromospheric activity and flares. Some studies like \citet{1998A&A...331..581D}, \citet{2003ApJ...583..451M} relate high activity with a faster rotation of the star. \citet{2019A&A...623A..44S} indicates that stronger magnetic fields in the active stars lead to Zeeman broadening of the individual lines in the band. H$\alpha$ is a good indicator of chromospheric activity. According to \citet{1979ApJ...234..579C}, H$\alpha$ goes into deeper absorption for low activity levels and into emission with increasing activity strength. Figure \ref{fig:result_spectra} shows H$\alpha$ emission specially visible in M stars. Stars in our sample which show a strongest H$\alpha$ and consequently a higher level of activity are the M4 Gaia DR2 1887754296868059648, the M8 2MASS J22512440+2952452 and M8 2MASS J08440874+7101007. 



We have compiled Gaia photometry (DR2) of our sample of M and L dwarfs. The absolute magnitude was estimated using Gaia parallaxes, reddening map for the interstellar extinction \citep{2011ApJ...737..103S}, and extinction coefficients for the Gaia photometric system \citep{2018MNRAS.479L.102C}. The Gaia database provides no valid parallax measurement for one star from our sample, 2MASS J03184214+0828002, thus the distance measured for its candidate companion {\citep{2012ApJ...760..152L}} was used for the calculations.

\begin{figure}[!t]
    \centering
    \includegraphics[width=0.8\linewidth]{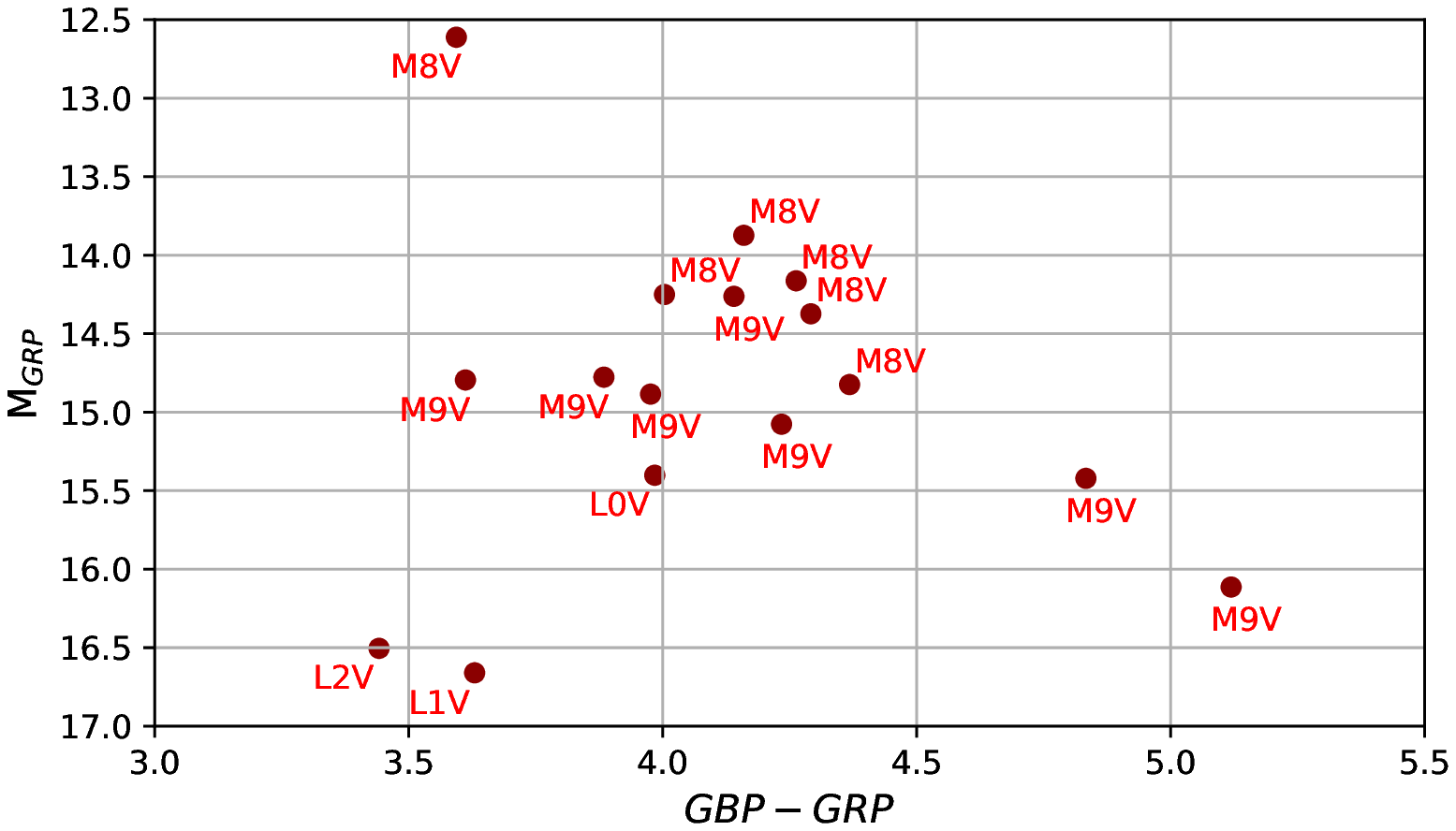}
    \caption{The Gaia color-magnitude diagram of our sample of stars. GBP and GRP stand for the blue and red Gaia bands respectively, while M$_{\textrm{GRP}}$ is the absolute magnitude in the red band. 
    }
    \label{fig:CMD2_Gaia}
\end{figure}

As it is shown in Figure \ref{fig:CMD2_Gaia}, the obtained color-magnitude diagram follows a trend in accordance with the expectations based on the Hertzsprung-Russel diagram. It can also be observed, that the different spectral classes occupy different slices on this diagram, although the applied classification was rather limited given the small amount of incorporated spectral templates in \texttt{pyhammer}. Our results proved that only a limited spectral classification can be done based on the Gaia photometry alone and it should be complemented by spectroscopy.

\section{Summary}

A spectral classification was carried out for 20 targets: 16 poorly studied M dwarfs, 3 L dwarfs and one K giant star. H$\alpha$ emission was detected in 6 objects (33$\%$ of the sample) with evidence of H$\alpha$ variability in one object. We have compared our spectral classification to Gaia photometry (DR2) concluding that spectral observations are needed for reliable spectral type determination.  

\acknowledgements
The authors would like to acknowledge support from ERASMUS+ grant number 2017-1-CZ01-KA203-035562, and the European Union's Horizon 2020 research and innovation programme under grant agreement No 730890 (OPTICON).

\bibliography{demo_caosp306}
\end{document}